# Weak lensing low multipoles

Albert Bonnefous[1]★ and Roya Mohayaee[1,2]★
[1]Sorbonne Université, CNRS, Institut d'Astrophysique de Paris, 98 bis Boulevard Arago, F-75014 Paris, France
[2]Rudolf Peierls Centre for Theoretical Physics, University of Oxford, Parks Road, Oxford OX1 3PU, UK

## ABSTRACT

We analyse the low-multipole components of the weak lensing convergence field in a Friedmann–Lemaître–Robertson–Walker (FLRW) universe. The low-multipole convergence field encodes the largest angle coherent potential gradients, essential for the assessment of large-angle features in data. To study this large-angle signal, we perform a combined analytical, numerical, and observational study. Starting from exact analytical expressions for the convergence power spectrum, we quantify how the dipole, quadrupole, and octupole build up with source redshift and show that, in $\Lambda$ cold dark matter ($\Lambda$CDM), they saturate at an amplitude of order $10^{-4}$. We then use full-sky, horizon-scale $N$-body simulations (QUIJOTE) to explore the dependence of this signal on the observer's environment, comparing random observers to 'Milky Way-like' observers. In parallel, we reconstruct the convergence field due to our local Universe with the 2MASS (Two Micron All-Sky Survey) Redshift Survey, with proper treatment of incompleteness and galaxy bias. We find that the observed low multipoles from observation are above the $\Lambda$CDM mean predictions, but in full agreement with Milky Way-like observers in the simulation. Finally, by converting the convergence dipole into a number count dipole, we test whether weak lensing can contribute to the cosmic dipole anomaly, an idea motivated by its natural alignment with the CMB (cosmic microwave background) dipole and by the fact that lensing, unlike clustering, cannot be removed by cross-matching surveys and thus survives in all high-redshift catalogues. We show that weak lensing by local structure contributes at most a few per cent to this observed anomaly.

**Key words:** gravitational lensing: weak – large-scale structure of Universe – cosmology: observations.

## 1 INTRODUCTION

Weak gravitational lensing (WL) probes the matter distribution in the Universe by tracing the cumulative deflection of light along the line of sight. Its main observable, the convergence field $\kappa(\boldsymbol{\theta})$, has been extensively studied on intermediate and small angular scales, accessible by observations and where effects such as cosmic shear are strongest (A. Refregier 2003). The reason why the largest angular scales of convergence – its dipole, quadrupole, and octopole – have remained mostly unexplored could be that the wide-angle surveys mapping the full or a good part of the sky have been quite rare so far. However, ongoing and upcoming wide-area surveys such as *SPHEREx* (*Spectro-Photometer for the History of the Universe, Epoch of Reionization, and Ices Explorer*; O. Doré et al. 2014), *Euclid* (Euclid Collaboration 2025), and Large Synoptic Survey Telescope/Vera C. Rubin Observatory (LSST/Rubin; Z. Ivezić et al. 2019) will provide us with an opportunity to study the wide-angle part of the spectrum. These surveys, which have large sky coverage, feature complex sky masks and non-uniform depth. In this context, a clean understanding of the low-$\ell$ convergence modes becomes unavoidable: they leak into masked-sky reconstructions, they perturb full-sky observables, and they contribute to wide-angle tests of the underlying cosmological model.

In a Friedmann–Lemaître–Robertson–Walker (FLRW) universe, it is expected that the lower multipoles of weak lensing convergence are dominated by the local superclusters, voids, and filaments within a few hundred megaparsecs – rather than by the distant homogeneous and isotropic Universe. The low multipoles quantify how the nearby matter distribution distorts the sky, and they are the only lensing modes to which full-sky, wide-angle observables are directly sensitive. Subsequently, the modelling of the low-$\ell$ convergence modes requires a careful treatment of both the nearby matter distribution and the specific environment of the observer. Studies of the effect of the local Universe on the convergence field and its low multipole has previously been done in R. Reischke et al. (2019), where they probe for the impact of the local overdensity on the angular power spectrum of convergence, and S. A. Ema et al. (2022) probe for the effect of the lower multipoles of convergence on cosmological parameter estimation. Lower multipole of the convergence field can also undermine the probe of primordial non-Gaussianity (PNG; D. Jeong, F. Schmidt & C. M. Hirata 2012). By evaluating and removing the convergence dipole, one can constrain $f_{\mathrm{NL}}$ (ruling out any primordial lensing dipole beyond cosmic variance) and avoid mistaking systematic or local-environment anisotropies for PNG signatures. Measuring the weak lensing dipole is also valuable when extracting the integrated Sachs–Wolf (ISW) effect. Careful modelling is required to disentangle genuine ISW correlations from spurious alignments caused by local density gradients and observer motion (see e.g. Y.-C. Cai et al. 2025).

★ E-mail: albert.bonnefous@iap.fr (AB); roya.mohayaee@physics.ox.ac.uk (RM)

In particular, characterizing the low-$\ell$ convergence field has implications for the *cosmic dipole anomaly*, using the so-called Ellis and Baldwin test (G. F. R. Ellis & J. E. Baldwin 1984), which relates our own velocity with the anisotropies in any light source number counts in the sky. This test has been carried out by N. J. Secrest et al. (2021) and N. Secrest et al. (2022, 2025), which showed a discrepancy between the number count dipole of distant light sources and the kinematic dipole inferred from the cosmic microwave background (CMB). This anomaly has not been explained ever since, and could have a huge impact on fundamental properties of the Universe (G. Domènech et al. 2022). However, weak lensing perturbs both solid angles and fluxes, producing an additional dipole in this number count even after nearby galaxies have been removed. Since this effect is created by the local matter distribution, the resulting dipole is expected to be closely aligned with both the clustering dipole and the CMB dipole. The impact of weak lensing on this number count dipole has been studied by C. Murray (2022), but the method, which did not evaluate the convergence explicitly, only provided an upper bound of $3 \times 10^{-2}$ on the convergence dipole, which is insufficient to rule out the weak lensing signal as a part of the cosmic dipole anomaly. Moreover, the method only probed for the dipole of convergence, which is insufficient to exclude weak lensing as a possible perturbation in carrying out this test, once we take into account the leakage from the higher multipoles due to masked sky.

Hereafter, we present a quantitative and detailed evaluation of the low multipoles of the convergence field within the $\Lambda$ cold dark matter ($\Lambda$CDM) framework:

(i) Theoretically, in Section 2, by deriving the exact full-sky expressions for $C_1^\kappa$, $C_2^\kappa$, and $C_3^\kappa$, without the Limber approximation and computing them for light sources at cosmological distances.

(ii) Numerically, in Section 3, by analysing the QUIJOTE simulations (F. Villaescusa-Navarro et al. 2020) and, crucially, placing observers in three distinct environments: random, Virgo-like, and Milky Way-like.

(iii) Observationally, in Section 4, by reconstructing the convergence multipoles from the full-sky 2MASS (Two Micron All-Sky Survey) Redshift Survey (2MRS) catalogue, and the distances computed by COSMICFLOWS-4 (CF4).

We then use our findings to estimate the impact of the low multipoles of weak lensing on the Ellis and Baldwin test in Section 5, and finally, using a simple toy model and estimations, we probe for the potential impact of structure beyond 2MRS in Section 6. Last, in Section 7, we rapidly explore how a deviation from an FLRW metric can modify the convergence field.

## 2 THEORETICAL ANALYSIS OF THE CONVERGENCE MULTIPOLES

### 2.1 Spherical harmonic analysis of the convergence field

The convergence field $\kappa(r_\mathrm{s}, \boldsymbol{\theta})$, defined along a line-of-sight direction $\boldsymbol{\theta}$ at comoving distance $r_\mathrm{s}$, can be expressed as (S. Dodelson & F. Schmidt 2020)

$$\kappa(r_\mathrm{s}, \boldsymbol{\theta}) = \int_0^{r_\mathrm{s}} \mathrm{d}r\, w_\kappa(r, r_\mathrm{s})\, \delta_\mathrm{m}(r, \boldsymbol{\theta}), \tag{1}$$

where the weight function $w_\kappa(r, r_\mathrm{s})$ in an FLRW universe is given by

$$w_\kappa(r, r_\mathrm{s}) = \frac{3H_0^2\Omega_\mathrm{m}}{2c^2}\frac{r(r_\mathrm{s}-r)}{r_\mathrm{s}\, a(r)}. \tag{2}$$

To quantify the angular structure of the convergence field, we expand it in spherical harmonics:

$$\begin{aligned} a_{\ell\mathrm{m}} &= \int \mathrm{d}\Omega\, Y^*_{\ell\mathrm{m}}(\boldsymbol{\theta})\, \kappa(r_\mathrm{s}, \boldsymbol{\theta}) \\ &= \int_0^{r_\mathrm{s}} \mathrm{d}r\, w_\kappa(r, r_\mathrm{s}) \int \mathrm{d}\Omega\, Y^*_{\ell\mathrm{m}}(\boldsymbol{\theta})\, \delta_\mathrm{m}(r, \boldsymbol{\theta}). \end{aligned} \tag{3}$$

The angular power spectrum of the convergence field is defined as

$$C_\ell^\kappa = \sum_m |a_{\ell m}|^2. \tag{4}$$

Of particular interest is the dipole amplitude, given by

$$\mathcal{D}_\kappa^2 = \frac{9C_1^\kappa}{4\pi}, \tag{5}$$

such that $\kappa(r_\mathrm{s}, \boldsymbol{\theta}) = \bar{\kappa} + \boldsymbol{\mathcal{D}}_\kappa \cdot \boldsymbol{\theta} + \mathcal{O}(l > 1)$, with $\bar{\kappa}$ is the mean value of convergence, and $\mathcal{O}(l > 1)$ represents the higher multipoles. Assuming linear growth of structure, the matter contrast evolves as $\delta_\mathrm{m}(r, \boldsymbol{\theta}) = D_+(r)\,\delta_\mathrm{m}(z = 0, r, \boldsymbol{\theta})$, where $D_+(r)$ is the linear growth factor. Using the results of Appendix A, the angular integral in equation (3) can be expressed in Fourier space, yielding the angular power spectrum:

$$\begin{cases} \langle C_\ell^\kappa \rangle = \frac{2}{\pi}\int_0^\infty \mathrm{d}k\, k^2\, P(k, z=0)\, W_\ell^\kappa(k, r_\mathrm{s})^2, \\ W_\ell^\kappa(k, r_\mathrm{s}) = \int_0^{r_\mathrm{s}} \mathrm{d}r\, w_\kappa(r, r_\mathrm{s})\, D_+(r)\, j_\ell(kr), \end{cases} \tag{6}$$

where $P(k, z = 0)$ is the present-day linear matter power spectrum, and $j_\ell$ are spherical Bessel functions.

Although this expression can be simplified using the Limber approximation (P. Lemos, A. Challinor & G. Efstathiou 2017), we avoid doing so here, since this approximation is accurate only for $\ell \gg 1$, even though it yields the correct 'order of magnitude' even at $\ell = 1$, the precise expression is more appropriate for our analysis (see Appendix C).

### 2.2 Interpretation of the convergence multipole profile

We now evaluate equation (6) numerically to determine how the lowest multipoles of the convergence field evolve with source redshift $z_\mathrm{s}$. The results are shown in Fig. 1. Our main finding is that the amplitude of the convergence low multipoles lies in the range $10^{-5}$–$10^{-4}$ and grows steadily with redshift up until $z_\mathrm{s} \simeq 0.2$ and then exhibits a strong reduction in its growth before saturating at around $z_\mathrm{s} \sim 1$. The quadrupole shows a similar evolution, with a mild increase followed by a saturation at a slightly higher redshift. The octopole and other low-$\ell$ modes behave analogously. Thus, the full calculation reveals a clear and physically meaningful trend: all low-$\ell$ weak lensing multipoles reach a maximum amplitude at intermediate redshifts and do not grow indefinitely with $z_\mathrm{s}$.

To understand the origin of this behaviour, we analyse the lensing weight function $w_\kappa(r, r_\mathrm{s})$ for different source redshifts (Fig. 2). Two key insights emerge. First, the lensing contribu-tion of a structure at a fixed comoving distance only increases with the light source redshift but eventually saturates once the source lies far beyond the lens. Second, weak lensing is intrinsically cumulative: it integrates the contribution of all intervening matter along the line of sight. One might therefore expect the convergence field as a whole to grow monotonically with $z_s$. However, this expectation is partially incorrect for the lowest multipoles; although the *total* convergence power increases with redshift, the angular structure of $\kappa$ prevents the dipole and quadrupole from growing indefinitely. The dipole probes only the very largest angular scales; structures at increasing distance spread on smaller angles on the sky and thus shift their contribution to higher multipoles. As a result, the low-$\ell$ components saturate once the dominant structures responsible for large-scale variations fall below the dipole and quadrupole angular scale. Thus, within $\Lambda$CDM, the weak lensing dipole and other low-$\ell$ multipoles flatten beyond $z_s \sim 1$ because distant structures project to smaller angular scales and contribute primarily to higher-$\ell$ modes. This behaviour is expected in any cosmological model that obeys large-scale homogeneity and isotropy and in which matter fluctuations diminish on the largest scales.

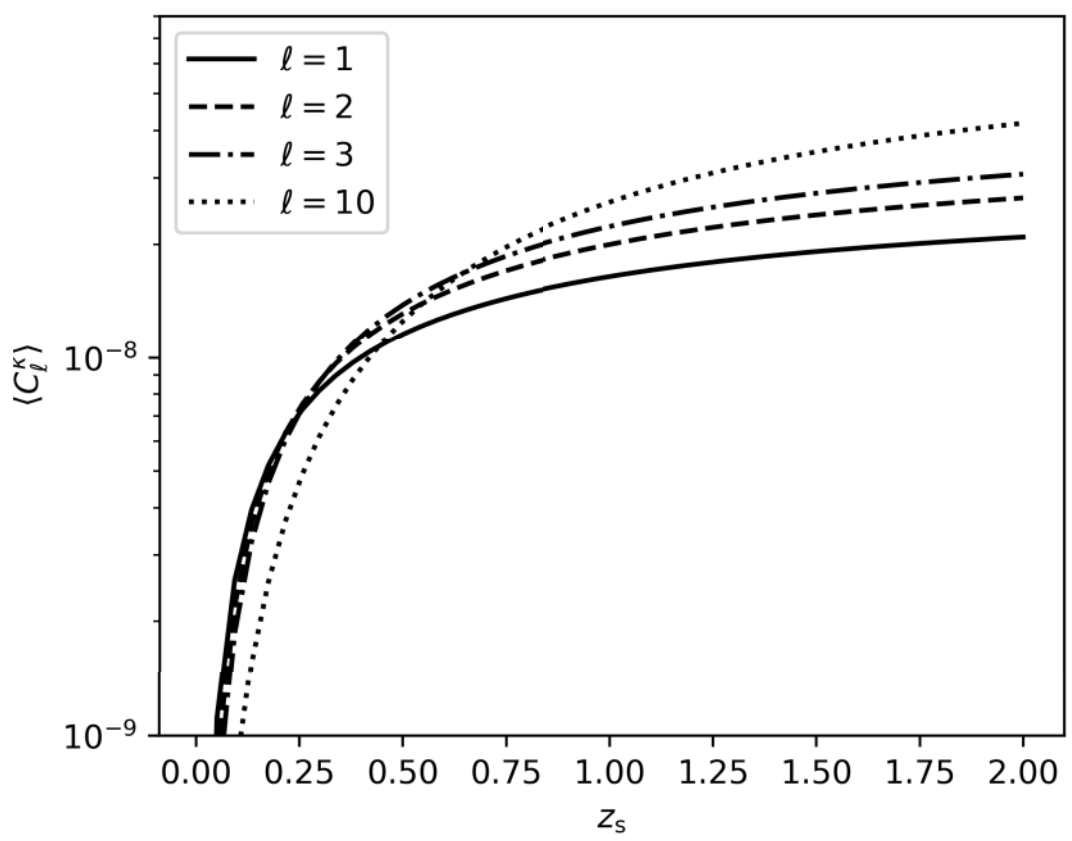

**Figure 1.** Evolution of the expected dipole ($\ell = 1$), quadrupole ($\ell = 2$), octopole ($\ell = 3$), and $\ell = 10$ contribution to the convergence power spectrum as a function of source redshift $z_s$, computed within $\Lambda$CDM. All low-$\ell$ modes exhibit a clear saturation at $z_s \simeq 1$–1.5.

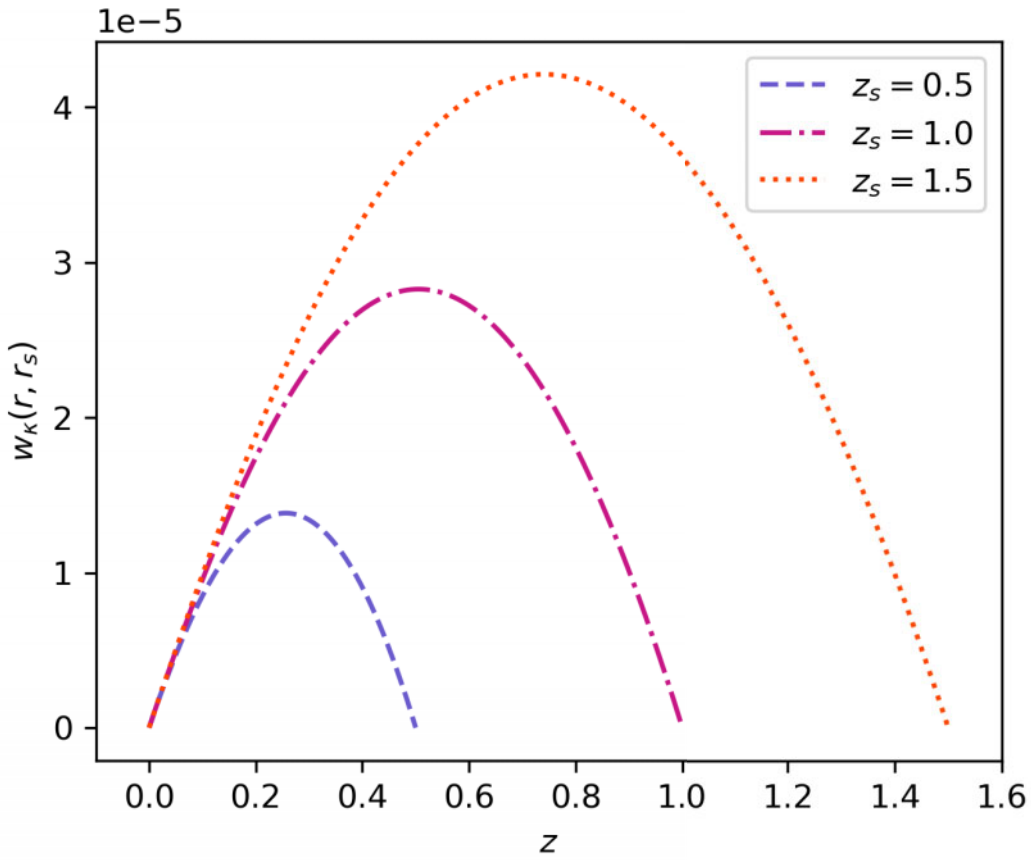

**Figure 2.** The weight function $w_\kappa(r, r_s)$ for sources at $z_s = 0.5$, 1.0, and 1.5. Although the peak of $w_\kappa$ lies roughly halfway between observer and source, this does *not* imply that structures at this distance dominate the low multipoles. Projection effects suppress the contribution of distant structures to the largest angular scales.

**Table 1.** Fitting function parameters for the low multipoles of the $\Lambda$CDM expected power spectrum of the convergence $\kappa$, between $z_s = 0$ and $z_s = 2$, along with the square root of the mean-squared deviation $\sqrt{\sigma^2}$ between the power spectrum and the fit.

| | $A$ | $p_1$ | $p_2$ | $z_0$ | $\sqrt{\sigma^2}$ |
|---|---|---|---|---|---|
| $\langle C_1^\kappa \rangle$ | $1.95 \times 10^{-8}$ | 1.49 | 1.28 | 0.26 | $7.4 \times 10^{-11}$ |
| $\langle C_2^\kappa \rangle$ | $2.54 \times 10^{-8}$ | 1.50 | 1.28 | 0.36 | $6.9 \times 10^{-11}$ |
| $\langle C_3^\kappa \rangle$ | $2.98 \times 10^{-8}$ | 1.51 | 1.28 | 0.43 | $7.5 \times 10^{-11}$ |

The expected angular power spectrum amplitude of the convergence field plotted in Fig. 1 can be approximated using a fitting function of the form

$$\langle C_\ell^\kappa \rangle (z_s) \approx A \frac{z_s^{p_1}}{z_s^{p_2} + z_0^{p_2}}. \quad (7)$$

The corresponding parameters for the $\ell = 1$, 2, and 3, between $z_s = 0$ and $z_s = 2$, are available in Table 1, as well as the square root of the mean-squared deviation $\sqrt{\sigma^2}$, such that $\sigma^2 = \left\langle \left( \langle C_\ell^\kappa \rangle (z_s) - A \frac{z_s^{p_1}}{z_s^{p_2} + z_0^{p_2}} \right)^2 \right\rangle$.

## 3 SIMULATIONS: CONVERGENCE MULTIPOLE IN A $\Lambda$CDM UNIVERSE WITH QUIJOTE

While equation (6) gives the expected value of the convergence multipoles in the $\Lambda$CDM framework, it represents only an average over all possible realizations of a $\Lambda$CDM universe. In other words, it is the statistical mean over all possible observer positions in such a cosmology. Our own location does not necessarily coincide with this mean. We do not live at a random point in space, but in a Galaxy embedded in a rich environment close to massive structures, superclusters (O. G. Kashibadze, I. D. Karachentsev & V. E. Karachentseva 2020).

To quantify how the convergence dipole depends on the observer's environment, we turn to cosmological $N$-body simulations. We use one realization from the QUIJOTE suite developed by F. Villaescusa-Navarro et al. (2020). This simulation adopts cosmological parameters consistent with *Planck* and evolves a periodic comoving box of side length $1000\,h^{-1}$ Mpc, with $1.68 \times 10^7$ dark matter particles with masses of $5.2 \times 10^{12}\,h^{-1}\,\mathrm{M}_\odot$ and a corresponding halo catalogue containing $3.5 \times 10^4$ haloes at redshift $z = 0$. Within this simulation, we define three classes of observers:

(i) *Random observers*, located at positions drawn uniformly within the simulation volume (i.e. not tied to any specific halo).

(ii) *Milky Way-like observers*, associated with particles located within $(12 \pm 4)\,h^{-1}$ Mpc of a Virgo-like halo and having peculiar velocities in the range $(370 \pm 150)\,\mathrm{km\,s^{-1}}$ with respect to the cosmological rest frame. In the simulation, 230 244 particles meet these requirements, which is 1.4 per cent of all particles.

(iii) *Virgo-like observers*, centred on haloes with masses in the range $(1.2 \pm 0.6) \times 10^{15}\,h^{-1}\,\mathrm{M}_\odot$. In our realization, 1506 haloes satisfy this criterion, which is 4.2 per cent of all haloes.

Motivated by the $\Lambda$CDM prediction that the convergence dipole sourced by nearby structure grows steadily up until $z \simeq$

0.2, and exhibits a strong reduction in its growth afterwards, we select 50 observers of each type and compute the convergence field up to this redshift. The number of observers is chosen such that their lines of sight probe largely distinct regions of the simulation volume. For each observer and for a sequence of redshifts, we pixelize the sky using HEALPIX[1] (K. M. Gorski et al. 2005) with 3072 pixels (13.4 $\mathrm{deg}^2$ each), and we generate the convergence field in each pixel with the knowledge of $\delta_{\mathrm{m}}$ and equation (1), and we measure the dipole, quadrupole, and octopole. The corresponding multipole amplitudes and their scatter are shown in Fig. 3. Error bars indicate the variance across the 50 realizations for a given observer class.

For observers chosen at random positions, the simulated multipoles closely follow the theoretical $\Lambda$CDM mean, particularly at the smallest redshifts. The modest deviations seen in the quadrupole and octopole can be attributed to the fact that we analyse a single simulation snapshot at $z = 0$, rather than constructing full light-cone realizations that include the explicit time evolution of the density field.

In contrast, observers living in Milky Way-like environments exhibit substantially enhanced multipole amplitudes at low redshift. For $z \lesssim 0.05$, *the dipole, quadrupole, and octopole can be up to an order of magnitude larger than the statistical mean*. This enhancement reflects the strongly anisotropic local environment of such observers, like ourselves, who reside near a massive Virgo-like cluster. To analyse the extent of the effect of environment, we can look at the extreme case of an observer living in a dense region of a *Virgo-like* cluster shows an even bigger departure from the $\Lambda$CDM expectations at these redshifts.

Despite these differences at very low redshift, by $z \simeq 0.1$ all observer classes converge to similar dipole amplitudes, in the range 2–4 $\times$ $10^{-5}$. The same trend is observed for the quadrupole and octopole. In other words, the impact of the local environment on the weak lensing multipoles is pronounced only in the very nearby Universe and rapidly diminishes with distance.

In interpreting our results, two methodological limitations should be kept in mind. First, our analysis is based on a single simulation snapshot at $z = 0$ rather than on full light-cone outputs. As a consequence, the redshift evolution of the matter distribution is not modelled directly: distant structures are sampled using the density field at the present epoch. This approximation is adequate for very shallow depths ($z \lesssim 0.1$), where cosmic evolution is minimal, but it may introduce biases in the multipoles once higher redshifts are considered. Second, although we select 50 observers of each type so that their lines of sight probe largely distinct regions of the simulation, true statistical independence cannot be achieved in a finite-volume box. Some overlap between the probed volumes is therefore unavoidable, particularly for distant shells.

These results support the expectation that, beyond $z \sim 0.2$, the convergence dipole should not deviate strongly from the $\Lambda$CDM ensemble average, even for observers in relatively special environments such as the Local Group. Whether the effect of local structure becomes entirely negligible at higher redshift in the context of $\Lambda$CDM remains an open question; addressing it will require larger-volume simulations with light-cone outputs extending to $z \gtrsim 1$ and full-sky catalogues going well beyond 2MRS. *SPHEREx* will offer an exciting opportunity to take this problem further (O. Doré et al. 2014).

[1] https://healpix.sourceforge.io/

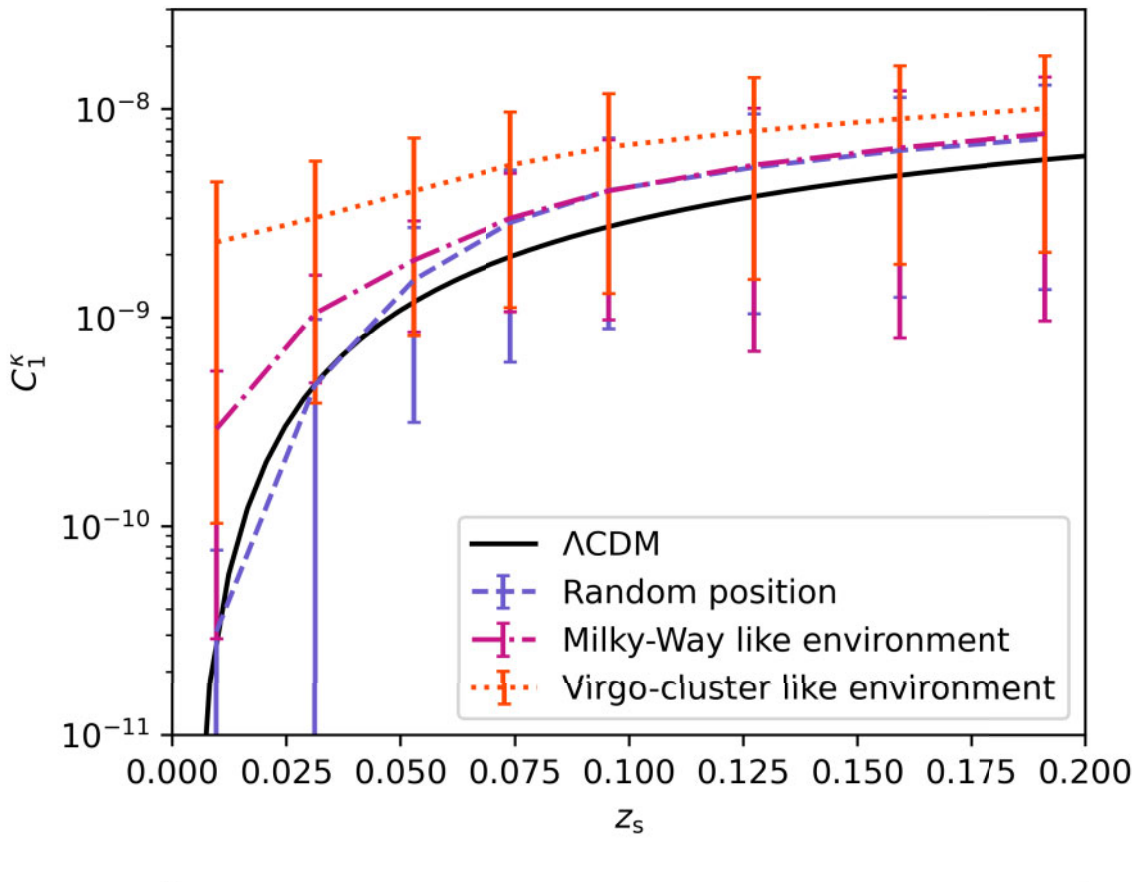

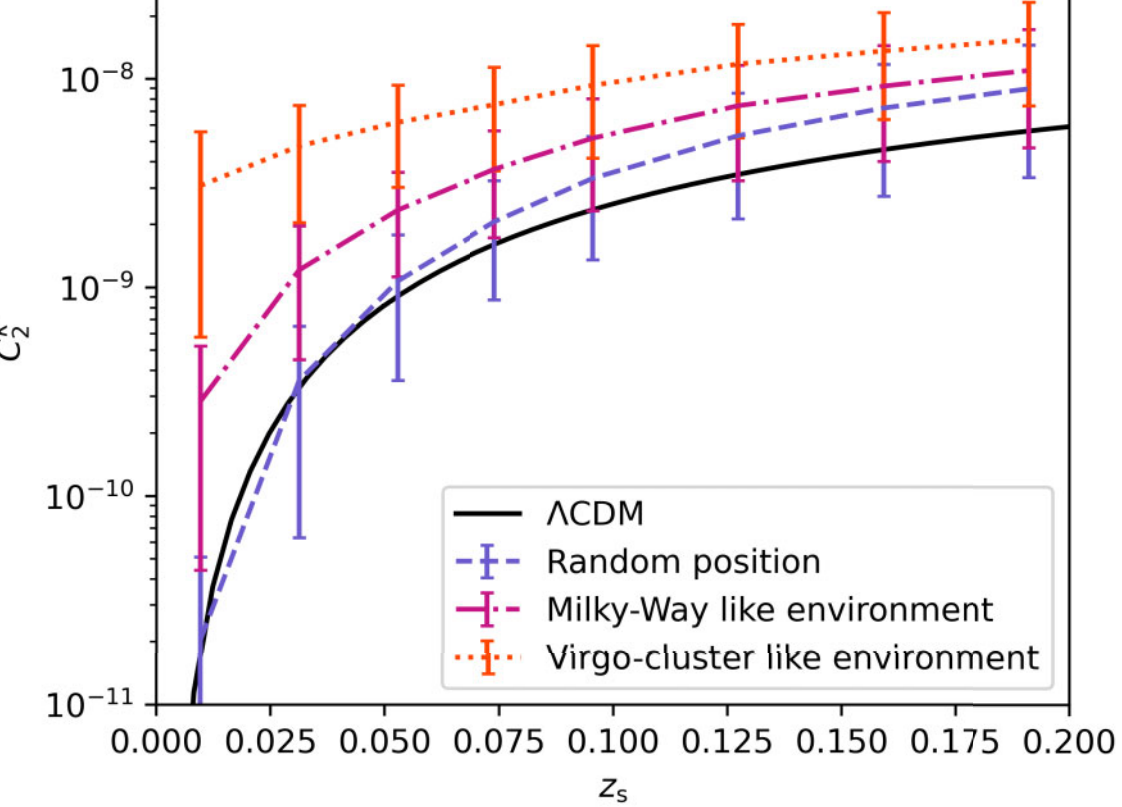

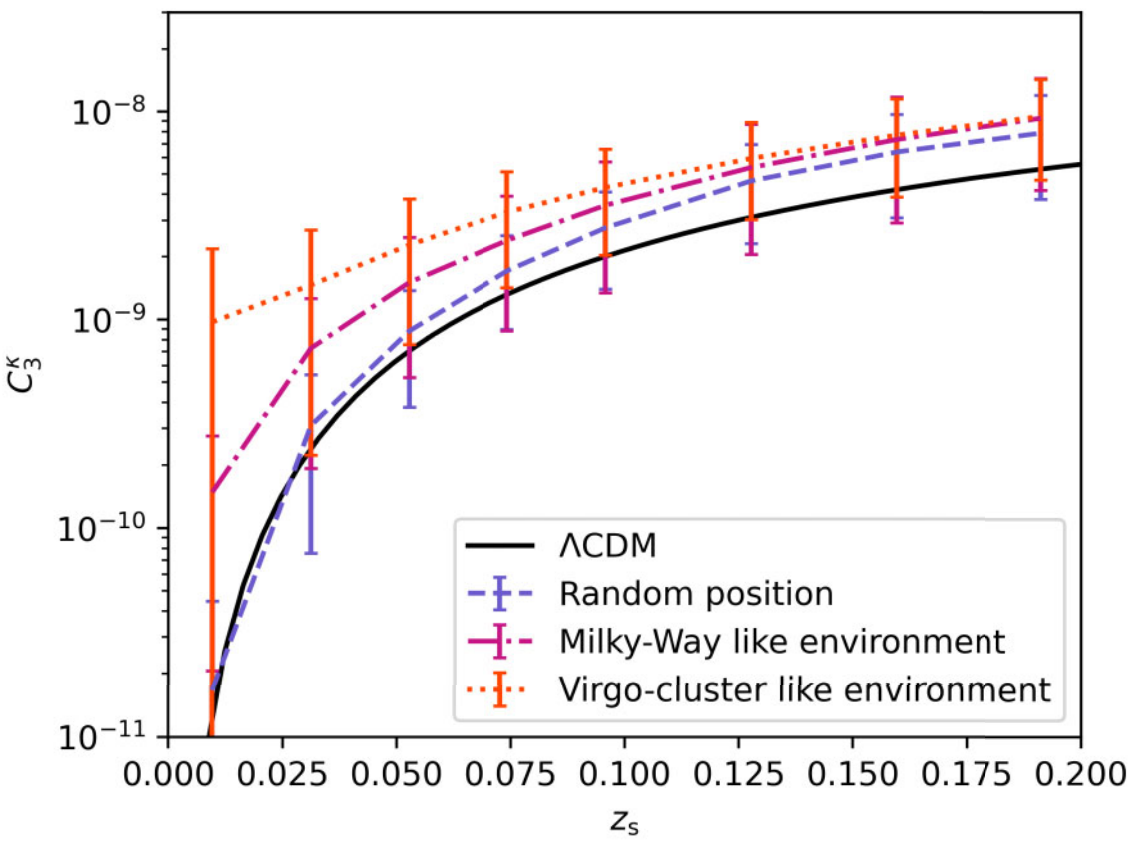

**Figure 3.** Dipole amplitude $C_1^\kappa$ (top), quadrupole amplitude $C_2^\kappa$ (middle), and octopole amplitude $C_3^\kappa$ (bottom) of the convergence field for the three classes of observers in the QUIJOTE simulation. For each observer type, 50 independent realizations are analysed. Error bars represent the variance across observers.

## 4 OBSERVATIONAL APPROACH: ESTIMATING THE CONVERGENCE MULTIPOLES WITH 2MRS

In this section, we aim to obtain a direct observational estimate of the convergence dipole, quadrupole, and octopole induced by the matter distribution in the nearby Universe. Because the low multipoles probe variations on the largest angular scales, they are particularly sensitive to sky coverage; even small sky cuts can introduce significant leakage. We therefore require a wide-area galaxy survey. For this purpose, we use the 2MRS (J. P. Huchra et al. 2012), a nearly full-sky compilation of more than 40 000 galax-ies. We use the extinction correlated $K_s$ magnitude of this survey; this band is located in the near-infrared wavelengths, around 2 μm, and mostly unaffected by extinction and stellar confusion (J. P. Huchra et al. 2012). Although the catalogue contains objects up to $z \simeq 0.2$, the survey is complete only up to $K_s = 11.75$. Our aim is therefore not to reconstruct a high-resolution convergence map, but to measure the large-scale multipoles of $\kappa(r_s, \boldsymbol{\theta})$ generated by the structures in our neighbourhood. We also use the distances for these galaxies given by CF4 (R. B. Tully et al. 2023), which compiles calculated distances for more than 55 000 close-by galaxies using different numerical methods. With this, we can compare the results obtained by neglecting the galaxies' peculiar velocity using redshift as a distance measurement with these calculated distances.

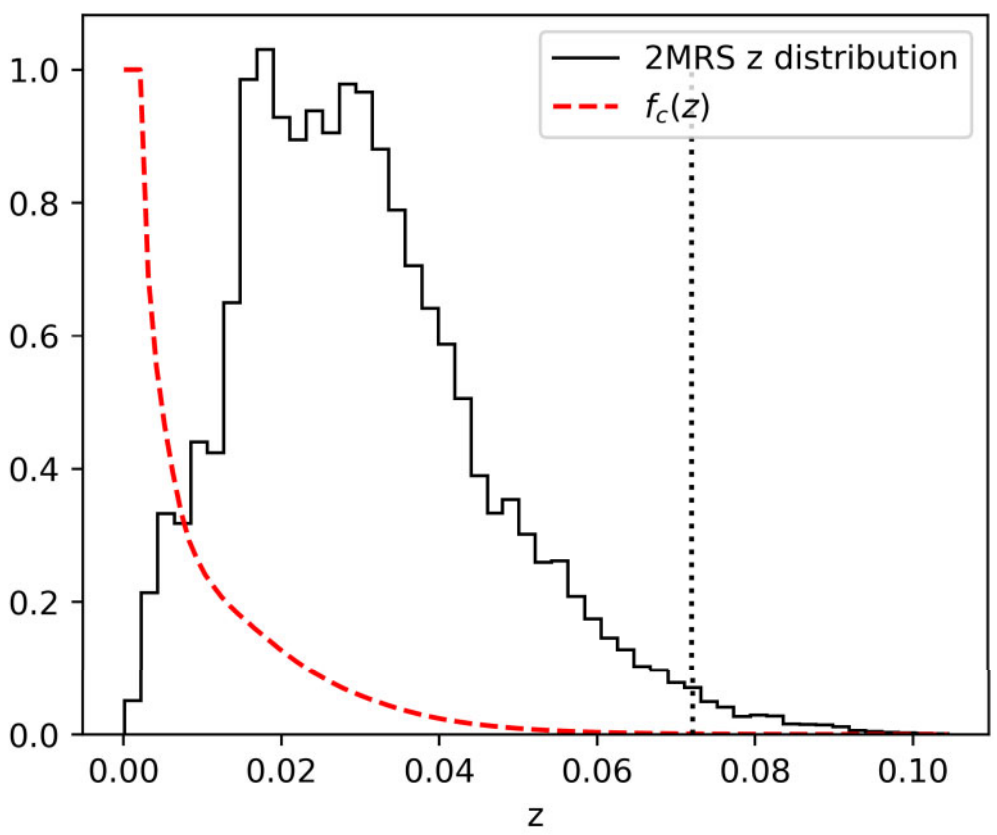

**Figure 4.** Completeness function $f_c(z)$ of the 2MRS survey after our cuts, along with the redshift distribution of the 42 580 surviving galaxies. The vertical dotted line marks the redshift at which the completeness falls below 0.1 per cent. We take this strict limit to achieve a higher precision of the convergence map.

Given the observed galaxy positions $\boldsymbol{r}_i$ and the linear bias relation $\delta_g = b_g\,\delta_m$ with $b_g \simeq 1.3$ for 2MRS (B. Pandey 2017), we construct the galaxy density contrast $\delta_g = n_g/\bar{n}_g - 1$, where $n_g(\boldsymbol{r}) = \sum_i \delta(\boldsymbol{r} - \boldsymbol{r}_i)$. To obtain an unbiased estimate of the underlying matter field, we correct for the redshift-dependent selection function caused by the fixed magnitude limit. Following C. Saulder et al. (2016), we model the completeness function $f_c(z)$ as the fraction of galaxies observable at redshift $z$ given the $K_s$-band luminosity function $\Phi(m)$:

$$f_c(z) = \frac{\int_{-\infty}^{-5\log_{10}(D_L(z))+m_{\rm limit}+5} \mathrm{d}m\,\Phi(m)}{\int_{-\infty}^{M_{\rm abs,min}} \mathrm{d}m\,\Phi(m)}. \tag{8}$$

Since the 2MRS scanning pattern is uniform and we use interstellar extinction-corrected magnitude here, the completeness function can be defined to be directionally independent. We apply several cuts. Regions with Galactic latitude $|b| < 10^\circ$ are masked; galaxies with negative redshifts are removed; we compute $K$-corrected absolute magnitudes, exclude objects with an apparent magnitude $K_s > 11.75$ to ensure uniform selection, and discard galaxies fainter than $M_K = -18$. After all cuts, 42 580 galaxies remain, for which 31 828 of them we have a distance given by CF4. The resulting completeness function is shown in Fig. 4. The completeness falls below 0.1 per cent at $z = 0.07$, beyond which the catalogue is no longer reliable for our purpose. This threshold is arbitrary and could have been chosen higher. Nevertheless, we compute convergence multipoles up to $z = 0.1$, noting that the highest redshift contributions are based on increasingly incomplete data.

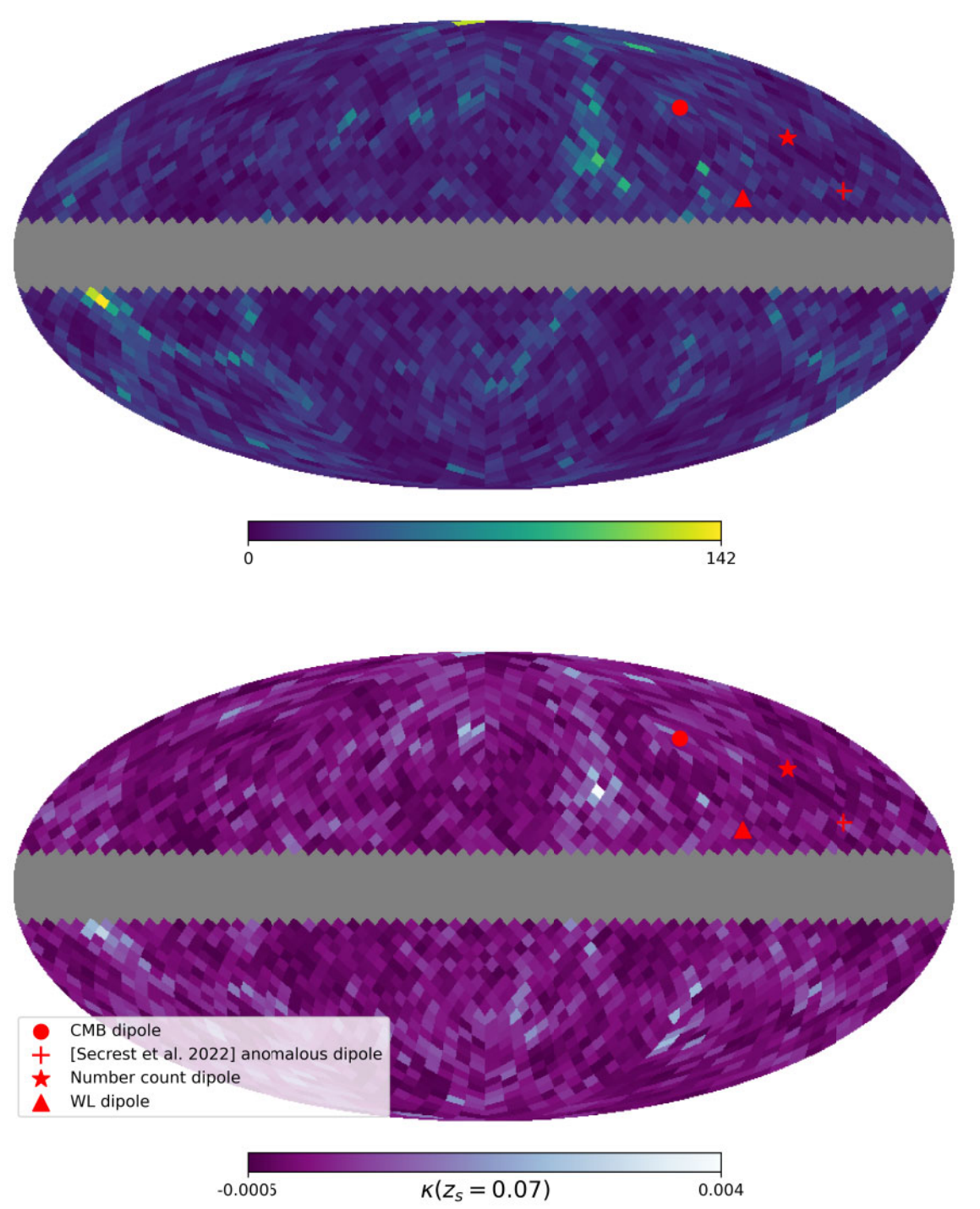

**Figure 5.** Top: Number of 2MRS sources per HEALPIX pixel. Bottom: Convergence map from 2MRS, with dipole directions overlaid: weak lensing dipole at $z_s = 0.07$, CMB dipole (Planck Collaboration I 2020), the 2MRS number count dipole, here dominated by clustering, and the number count dipole anomaly obtained in N. Secrest et al. (2022). Grey region: Galactic mask ($|b| < 10^\circ$).

With the completeness correction included, the matter overdensity becomes

$$\delta_m(\boldsymbol{r}) = \frac{1}{b_g}\left(\frac{n_{\rm 2MRS}(\boldsymbol{r})}{\bar{n}_{\rm 2MRS}\,f_c(r)} - 1\right). \tag{9}$$

In the same way as with the simulation, we pixelize the sky using HEALPIX, with 3072 pixels (13.4 deg$^2$ each), and compute the convergence field $\kappa(r_s, \boldsymbol{\theta})$ with the knowledge of $\delta_m$ in each pixel, and extract its multipoles. The dipole direction obtained at $z = 0.07$ is shown in Fig. 5. The convergence dipole lies roughly 39° from the CMB dipole (Planck Collaboration I 2020) and 34° from the 2MRS galaxy number count dipole, which is dominated by clustering here, the intrinsic anisotropies in the repartition of galaxies due to structure formation, see Appendix B. The convergence of low-multipole amplitudes reconstructed from 2MRS using distances from CF4 and compared with the 2MRS redshift is shown in Fig. 6. Both methods give similar results, showing that these low multipoles are not very sensitive to the peculiar velocities of each galaxy. These multipoles stay persistently above the ΛCDM prediction, but are perfectly consistent with Milky Way-like observations in the QUIJOTE simulation.

Several limitations of our observational reconstruction should be acknowledged. First, the 2MRS completeness drops rapidly beyond $z \simeq 0.07$, and although we compute the convergence multipoles up to $z = 0.1$, the highest redshift contribution is based on increasingly incomplete data. Second, we assume a constant linear galaxy bias for all 2MRS galaxies; this approximation is reasonable at the shallow depths considered here, but may introduce small systematic shifts in the inferred density field. Finally, our sky mask removes the Galactic plane, which inevitably introduces some degree of multipole leakage despite the large sky fraction. These caveats do not impact our main conclusions but should be kept in mind when interpreting the precise multipole amplitudes.

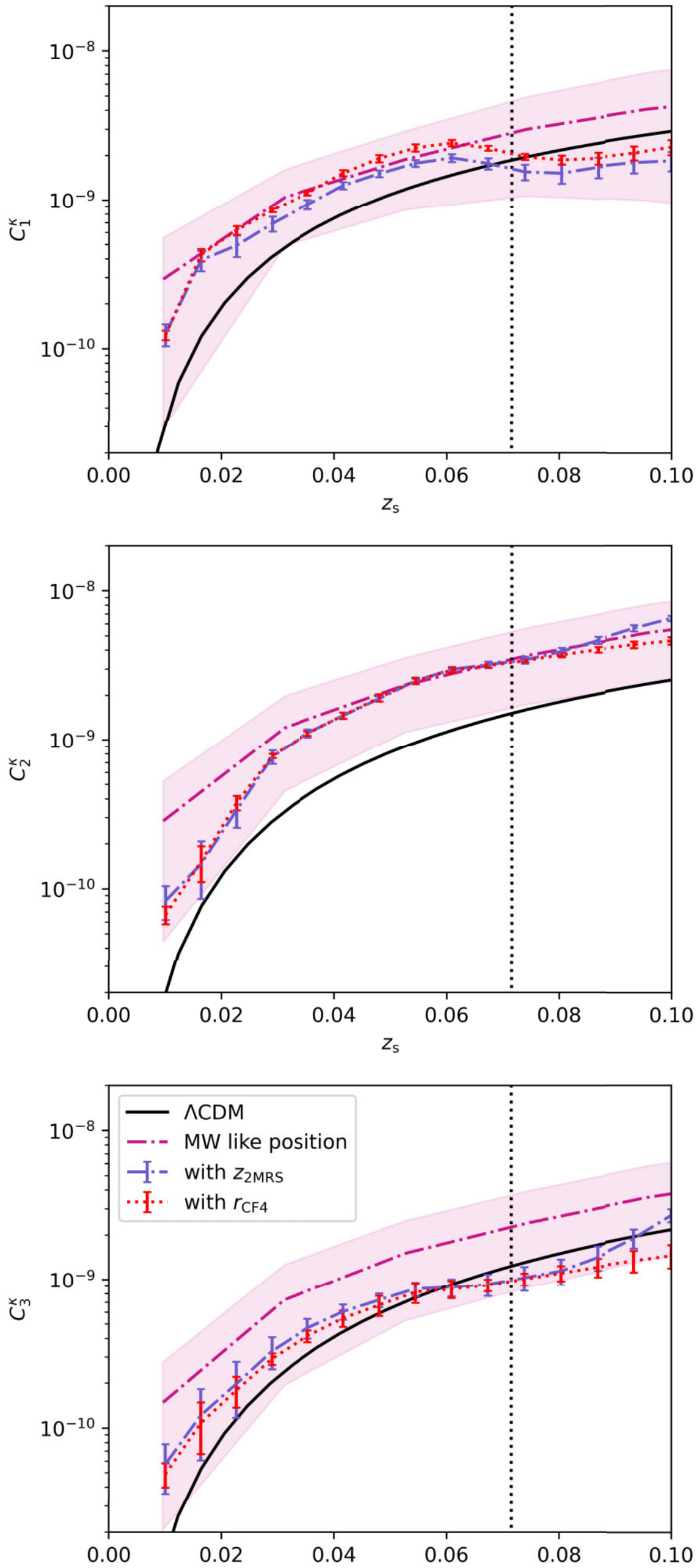

**Figure 6.** Comparison of the dipole amplitude $C_1^\kappa$ (top), quadrupole amplitude $C_2^\kappa$ (middle), and octopole amplitude $C_3^\kappa$ (bottom), calculated with the 2MRS survey, using distances $r_{\mathrm{CF4}}$ calculated with CF4, and distances given by a direct conversion of the 2MRS redshifts $z_{\mathrm{2MRS}}$. In black, the expected results within $\Lambda$CDM, and the vertical dotted line marks the redshift $z = 0.07$ at which the 2MRS completeness falls below $10^{-3}$. All these multipoles are compared with the results of the simulation with Milky Way-like observations.

## 5 WEAK LENSING CONTRIBUTION TO THE COSMIC DIPOLE ANOMALY

What is the motivation behind using weak lensing to explain the cosmic dipole anomaly? First, weak lensing convergence is induced by local structures and the CMB dipole point in approximately the same direction. A second motivation is that, although the use of quasars and radio galaxies minimizes contamination from the clustering dipole, the weak lensing signal persists in all cases. Even when a low-redshift cut-off is applied, the weak lensing contribution is not removed. Therefore, without a reliable estimate of its magnitude, one cannot properly subtract the weak lensing convergence from the observed cosmic dipole anomaly. For these reasons, we next study the contribution of weak lensing convergence to the cosmic dipole anomaly.

### 5.1 Number count dipole

Here, we derive the expression for the number count dipole by following the approach of G. F. R. Ellis & J. E. Baldwin (1984), while extending it to include the effects of WL on both angular distortion and frequency shift. The above expression was first used in C. Murray (2022). In this work, the weak lensing convergence itself was not directly measured. Instead, an upper limit on its possible contribution was put by making two conservative assumptions: first, they attributed all remaining systematic or ‘nuisance’ contributions to the observed number count dipole to weak lensing convergence; second, they assumed that the convergence field was a pure dipole completely aligned with the CMB. Under this maximal-assumption scenario, an upper bound for convergence was obtained. In our work, we evaluated the weak lensing convergence itself, and after performing spherical harmonics, we obtained expressions for the low multipoles. We link the number count dipole to an underlying convergence field – but we go beyond C. Murray (2022) analysis by explicitly evaluating the convergence dipole itself.

We consider a population of sources at rest in the cosmological frame, located at a comoving distance $r_s$, corresponding to a redshift $z_s$. Weak lensing modifies observations, and therefore the source number count, along a given direction $\boldsymbol{\theta}$ through the convergence parameter $\kappa$, which is given by equations (1) and (2). Since we are concerned only with the angular position of sources on the sky, the convergence $\kappa$ is the relevant weak lensing observable; we do not consider shear, which affects image shapes.

So far, we have evaluated the convergence field for an observer at rest in the cosmic rest frame. If the observer has a motion – which is the case for the Sun or the Local Group – then this motion affects the weak lensing signal. Taking into account both the observer’s motion and weak lensing, the observed solid angle transforms as

$$\mathrm{d}\Omega_{\mathrm{obs}} = \mathrm{d}\Omega_{\mathrm{rest}}\, \delta(\boldsymbol{\theta})^2 \left(1 + 2\kappa(\boldsymbol{\theta})\right), \tag{10}$$

where $\mathrm{d}\Omega_{\mathrm{rest}}$ is the solid angle in the cosmological rest frame, $\mathrm{d}\Omega_{\mathrm{obs}}$ is the corresponding solid angle in the observer’s frame, and

$$\delta(\boldsymbol{\theta}) = 1 + \boldsymbol{\beta} \cdot \boldsymbol{\theta} + \mathcal{O}(\beta^2)$$

accounts for Doppler and aberration effects due to the observer’s velocity.

Assuming the sources' flux follows a power-law dependence on frequency, $F \propto \nu^{-\alpha}$, then at fixed observed frequency, the observed flux becomes

$$F_{\rm obs} = \delta(\boldsymbol{\theta})^{1+\alpha}\left(1 + 2\kappa(\boldsymbol{\theta})\right) F_{\rm rest}. \tag{11}$$

Due to instrumental sensitivity limits, sources below a certain flux threshold $F_{\rm min}$ remain undetectable. Suppose that for a monochromatic survey, the cumulative number of sources above this limit follows a power-law distribution:

$$\frac{{\rm d}N}{{\rm d}\Omega}(> F_{\rm min}) \propto F_{\rm min}^{-x}.$$

Note that only the faintest sources, which are close to the flux limit, are susceptible to appearing or disappearing with Doppler-boosting and magnification; therefore, both indices $x$ and $\alpha$ should be measured for these sources at the flux limit (S. Hausegger 2024). For more complicated spectra in photometric surveys where a power law cannot be fitted, it has also been shown that the Ellis and Baldwin formula holds, with a redefinition of the coefficient $x$ and $\alpha$ (A. Bonnefous 2026). This assumption about the power-law indices $\alpha$ and $x$ is reasonable, since only sources near the detection limit contribute significantly to the dipole. Moreover, this approximation is well justified for radio-selected active galactic nuclei and quasars, whose flux is dominated by non-thermal synchrotron emission and therefore follows a power-law spectrum. The observed number of sources per unit solid angle becomes

$$\frac{{\rm d}N}{{\rm d}\Omega_{\rm obs}}(> F_{\rm min}, \boldsymbol{\theta}) = \delta(\boldsymbol{\theta})^{2+x(1+\alpha)}\left(1 + 2\kappa(\boldsymbol{\theta})\right)^{x-1}\frac{{\rm d}N}{{\rm d}\Omega_{\rm rest}}.$$

To first order in $\beta$ and $\kappa$, this simplifies to

$$\frac{{\rm d}N}{{\rm d}\Omega_{\rm obs}}(> F_{\rm min}, \boldsymbol{\theta}) = \bar{N}\left(1 + \boldsymbol{\mathcal{D}}_{\rm kin}\cdot\boldsymbol{\theta} + 2(x-1)\kappa(\boldsymbol{\theta})\right), \tag{12}$$

where $\bar{N}$ is the mean number count, and the kinematic dipole is given by

$$\boldsymbol{\mathcal{D}}_{\rm kin} = \left(2 + x(1+\alpha)\right)\boldsymbol{\beta}. \tag{13}$$

This result may seem counterintuitive: since the convergence $\kappa$ is proportional to the matter overdensity $\delta_{\rm m}$, one might expect an increase in the number of observed sources in high-density regions. Indeed, weak lensing is often used to detect faint sources by magnifying their flux. However, when the power-law index satisfies $x < 1$, the number count actually decreases in regions of high matter density. This stems from the competing effects of lensing: magnification increases the flux and allows more sources to be detected, but simultaneously stretches the solid angle and thus dilutes the number density.

The net effect depends on the value of $x$. We focus on extracting the dipole component $\boldsymbol{\mathcal{D}}_{\rm WL}$ of the weak lensing contribution, defined through the expansion

$$2(x-1)\kappa(r_{\rm s}, \boldsymbol{\theta}) = m_{\rm WL} + \boldsymbol{\mathcal{D}}_{\rm WL}\cdot\boldsymbol{\theta} + \mathcal{O}(\ell > 1), \tag{14}$$

where $m_{\rm WL} = 2(x-1)\bar{\kappa}$ is the monopole (the sky average of the lensing term), and $\mathcal{O}(\ell > 1)$ denotes higher order multipoles. Since $x$ is survey dependent, we isolate the dipole of the convergence $\boldsymbol{\mathcal{D}}_{\kappa}$ and relate it to the lensing dipole via

$$\boldsymbol{\mathcal{D}}_{\rm WL} = 2(x-1)\boldsymbol{\mathcal{D}}_{\kappa}. \tag{15}$$

For example, in N. J. Secrest et al. (2021), with the CatWISE quasar catalogue $x = 1.89$ at the 0.078 mJy flux limit, the convergence dipole must be multiplied by 1.78 to estimate the weak lensing dipole. In contrast, in N. Secrest et al. (2022), for the NRAO VLA Sky Survey (NVSS) survey $x = 0.77$ at the flux density limit of 4 mJy, the factor becomes $-0.46$. These results suggest that while weak lensing cannot fully account for the observed dipole amplitude in either catalogue, it may help reconcile at least to some extent their differences – particularly the fact that the NVSS dipole (see e.g. J. Colin et al. 2017) is smaller than that of CatWISE (N. Secrest et al. 2022). However, looking at Fig. 7, we find that the corresponding dipole amplitude is completely negligible with respect to the number count dipole in these surveys. The weak lensing dipole for the CatWISE quasar is at most $2 \times 10^{-4}$ against a total number count dipole of $\mathcal{D}_{\rm N} = (1.48 \pm 0.16) \times 10^{-2}$, while for the NVSS radio galaxies, we have at most $6 \times 10^{-5}$ against $\mathcal{D}_{\rm N} = (1.23 \pm 0.25) \times 10^{-2}$.

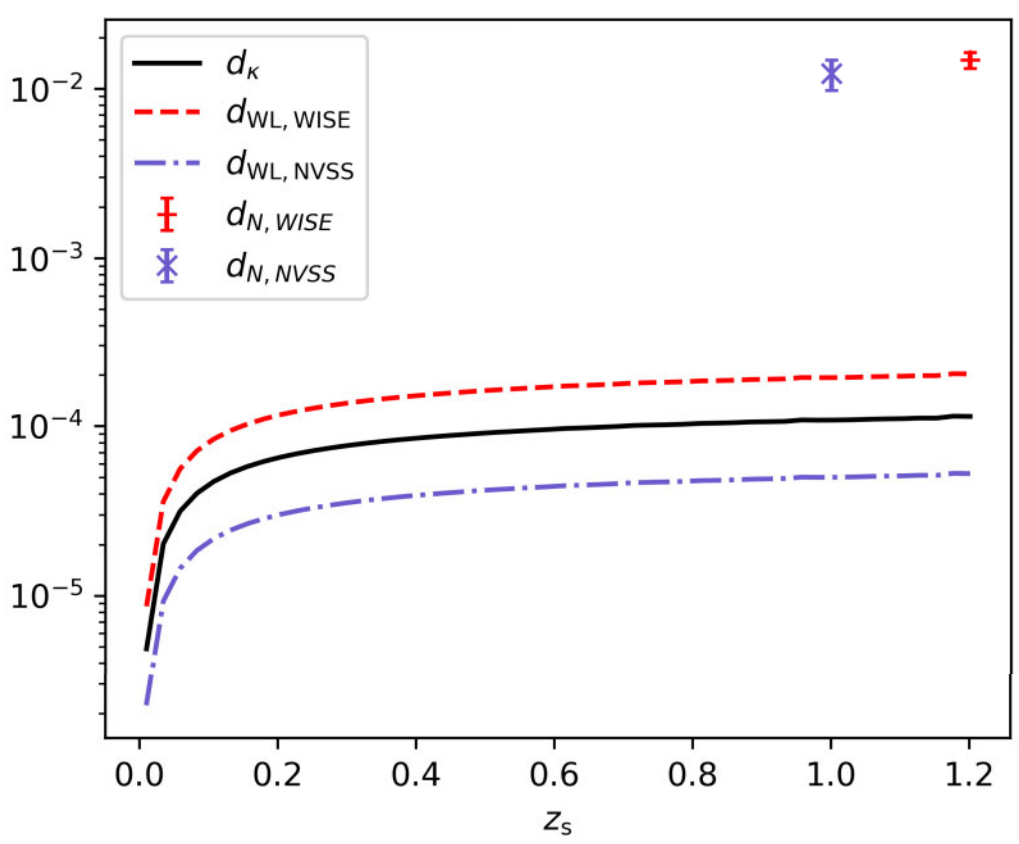

**Figure 7.** In black, integration of the convergence dipole up to a redshift of $z = 1.2$, using equation (6). Dashed lines are the corresponding weak lensing dipole for the CatWISE and NVSS catalogues, taking into account the values of $x$ for these catalogues from N. Secrest et al. (2022). Above, the corresponding number count dipoles for those two catalogues, at their corresponding mean redshifts. Within $\Lambda$CDM, the impact of weak lensing on these dipoles is negligible.

## 5.2 Leakage of higher multipole WL convergence into number count dipole

It is somehow counterintuitive that as we go to higher multipoles, the weak lensing convergence increases. Hence, an immediate question is to evaluate the contribution to the weak lensing number count dipole from higher order multipoles from leakage due to masked sky. To complete this analysis, it is necessary to estimate the impact of leakage. It is completely dependent on the coupling matrix between the real and estimated multipoles (A. Abghari et al. 2024; S. Hausegger et al. 2026):

$$M_{\ell m \ell' m'} = \int_{\rm obs} {\rm d}\Omega\, Y^{*}_{\ell' m'}(\boldsymbol{\theta}) Y_{\ell m}(\boldsymbol{\theta}), \tag{16}$$

where $\int_{\rm obs}$ is taken in the unmasked part of the sky $f_{\rm obs}$, such that we have

$$a^{\rm obs}_{\ell m} = \sum_{\ell' m'} M_{\ell m \ell' m'} a_{\ell' m'}. \tag{17}$$

And we can easily show that $|M_{\ell m \ell' m'}|^2 < 1$, so the leakage of one multipole into another is therefore at most of the order of the leaking multipole itself. Moreover, leakage mostly happens between close multipoles. Therefore, if the dipole itself can be neglected, so does the possible leakage of the quadrupole and the octopole onto it. Weak lensing can therefore be excluded as a possible explanation for the cosmic dipole anomaly.

## 6 BEYOND 2MRS: CAN STRUCTURES BEYOND 2MRS EXPLAIN THE COSMIC DIPOLE ANOMALY?

The 2MRS catalogue provides the most complete full-sky view of the nearby Universe, but it becomes highly incomplete beyond $z \simeq 0.07$–$0.08$. As a result, our reconstruction of the convergence multipoles based on 2MRS includes only the structures contained within this shallow volume. The weak lensing dipole obtains most of its amplitude from the very low redshift Universe, but massive overdensities lying just beyond the 2MRS completeness limit could, in principle, contribute to the low multipoles of the convergence field, especially for background sources at higher redshifts. An important question is therefore: how large could the contribution of structures beyond 2MRS be, and could those standing to other observational constraints, modify the weak lensing dipole enough to alleviate certain large-angle tensions, for example, the cosmic number count dipole anomaly? To address this question, we combine simple analytic arguments with numerical tests.

Several large structures lie just beyond the effective 2MRS horizon. For example, the Shapley Supercluster at $r_{\rm L} \simeq 150$ Mpc is one of the most massive concentrations in the nearby Universe, with a mass of order $\sim 10^{17}\,{\rm M}_\odot$ (see e.g. S. Raychaudhury 1989; D. Proust et al. 2006). Similarly, the Sloan Great Wall at $z \simeq 0.07$ has a mass $\sim 2.5 \times 10^{16}\,h^{-1}\,{\rm M}_\odot$ and an extent exceeding 200 Mpc (M. Einasto et al. 2016). Such structures generate measurable large-angle anisotropies in the local matter distribution and have been invoked in studies of bulk flows (R. Watkins, H. A. Feldman & M. J. Hudson 2009; G. Lavaux et al. 2010; J. Colin et al. 2011; U. Feindt et al. 2015).

Although these structures lie outside the region where 2MRS is complete, they could, in principle, contribute to the low multipoles of the convergence, since weak lensing integrates matter along the line of sight. The question is whether their impact could be large enough to matter at high source redshift ($z_s \sim 1$). Because the lensing kernel contains the factor $r(r_s - r)/r_s$, structures located at 150–300 Mpc experience only mild geometric suppression when $r_s \gg r$.

Let us consider a Shapley-like structure, a spherical overdensity of radius $R_{\rm S} \simeq 25$–$40$ Mpc and located at a comoving distance of about $r_{\rm L} \simeq 200$ Mpc and $\delta_{\rm m}^{\rm S} \simeq 3$–$5$, which subtends only a small fraction of the sky:

$$f_{\rm S} \sim \left(\frac{R_{\rm S}}{r_{\rm L}}\right)^2 \sim 10^{-2}.$$

For $z_s \gg z_{\rm L}$, the lensing efficiency saturates,

$$\frac{r_{\rm L}(r_s - r_{\rm L})}{r_s a(r_{\rm L})} \simeq r_{\rm L}, \tag{18}$$

so increasing the source redshift from $z_s = 0.2$ to $z_s = 1$ changes the efficiency by only $\sim$20–30 per cent. We can therefore make a simple approximation to see how this impacts the convergence dipole. The convergence field being linear in $\delta_{\rm m}$, the convergence change created this structure is of the order

$$\delta\kappa_{\rm S} \simeq \frac{3H_0^2\Omega_{\rm m}}{2c^2} R_{\rm S}\, r_{\rm L}\, \delta_{\rm m}^{\rm S} \simeq 10^{-3}. \tag{19}$$

Since the sky fraction is small, most of the contribution therefore projects into higher multipoles, and the dipole amplitude change can be expressed using the spherical decomposition of $\delta\kappa_{\rm S}$:

$$a_{1m} = \int_{f_{\rm S}} {\rm d}\Omega\, Y_{1m}^*(\boldsymbol{\theta})\delta\kappa_{\rm S} \simeq f_{\rm S}\, \delta\kappa_{\rm S}\, Y_{1m}^*(\boldsymbol{\theta}_{\rm S}), \tag{20}$$

where $\boldsymbol{\theta}_{\rm S}$ is the direction of the structure in the sky, and $\int_{f_{\rm S}}$ is taken over the part of the sky occupied by this structure. The convergence dipole $|\mathcal{D}_\kappa|^2 = \sum |a_{1m}^2| \times 9/4\pi$ is therefore modified *at most* proportionally to this sky fraction, and becomes at most

$$\delta\mathcal{D}_\kappa \simeq f_{\rm S}\, \delta\kappa_{\rm S}\, \frac{9}{4\pi} \simeq 10^{-5}. \tag{21}$$

This agrees with earlier assessments of the gravitational impact of Shapley on local flows, where its contribution is found to be significant but not dominant (U. Feindt et al. 2015). Even combining several known structures (e.g. Shapley Supercluster + Sloan Great Wall) would raise the estimate only modestly. The contributions add only partially in phase and remain at the level $\sim 10^{-6}$–$10^{-5}$. Even considering an $\mathcal{O}(10)$ of such structure in our vicinity, they would all need to be in the same broad direction in the sky, without any counterbalancing structure on the other part of the sky to maximize their impact on the dipole. This would be highly improbable.

However, for a bigger structure in the sky, we cannot make this approximation, so we construct a simple toy model. We take the 2MRS-derived density field out to $z = 0.07$ and fill the region beyond it with a uniform background plus a single, extremely massive spherical overdensity placed along the CMB dipole direction. This configuration is deliberately tuned to maximize the dipole, since the added structure is not compensated by any counterbalancing density in the opposite direction. This spherical overdensity has a mass of $M \sim 5 \times 10^{18}\,{\rm M}_\odot$, with a diameter of $\sim$400 Mpc. We place the overdensity at $z_{\rm L} = 0.1$, and we compute the convergence field for background sources at $z_s = 1.0$. The resulting dipole amplitudes are about $5 \times 10^{-4}$. The dipole increases only by a factor of a few, even with such extreme and unrealistically massive structures, whose mass exceed that of Shapley or the Sloan Great Wall (J. R. Gott et al. 2008; M. Einasto et al. 2016) by about two orders of magnitude and are highly unexpected in $\Lambda$CDM.

## 7 WEAK LENSING CONVERGENCE BEYOND FLRW BACKGROUND

In this work, we have computed the weak lensing convergence assuming a FLRW background space–time, for which the homogeneous and isotropic geometry implies that no lensing signal is generated at the background level. In this framework, the convergence arises solely from *cosmological perturbations*, in particular scalar density fluctuations, through Ricci and Weyl focusing (WL convergence and shear) along null geodesics. This assumption underlies the standard weak lensing formalism that we have used here and is well motivated within the $\Lambda$CDM cosmology.

However, the weak lensing formalism can be extended to more general cosmological backgrounds, including homogeneous but anisotropic metrics. In such cases, light propagation is better treated using the Sachs optical equations or, equivalently, the Jacobi map, rather than the standard isotropic lensing kernel (see e.g. J. Kristian & R. K. Sachs 1966). A general perturbative framework for cosmological perturbations has been developed by T. S. Pereira, C. Pitrou & J.-P. Uzan (2007), who showed that *background anisotropies* induce direction-dependent angular-diameter distances.

More generally, weak lensing convergence can be defined in an arbitrary space–time through the Jacobi map governing the evolution of infinitesimal light bundles along null geodesics, and an expression of the Jacobi matrix of weak lensing within a Bianchi I universe in particular can be found in P. Fleury, C. Pitrou & J.-P. Uzan (2015), which can then be decomposed in the more usual convergence and shear terms. However, in the most general case where the deviations from FLRW are small, as required by current observational constraints, a perturbative approach can be adopted in which the convergence is expanded around the FLRW result. At leading order in terms of the conformal shear $\sigma$ of the metric and in terms of perturbation of the metric due to the gravitational potential $\Phi$ and $\Psi$, to get an order-of-magnitude estimate, one may write a double expansion *schematically* (C. Pitrou, T. S. Pereira & J.-P. Uzan 2015):

$$\kappa(\hat{\boldsymbol{n}}, z) = \kappa_{\rm FLRW} + \kappa_{\rm pertFL} + \kappa_{\rm bg} + \kappa_{\rm pert/aniso}, \quad (22)$$

where $\kappa_{\rm FLRW}$ is the usual convergence term in a flat FLRW metric, $\kappa_{\rm pertFL}$ is the term arising from the perturbation of this metric, depending on the perturbed gravitational potential, the new term $\kappa_{\rm bg}$ denotes the contribution arising from the anisotropic background geometry itself, $\kappa_{\rm pertFL}$ describes the coupling between the anisotropy and the perturbed gravitational potential and all terms on the right-hand side have the argument $(\hat{\boldsymbol{n}}, z)$. However, within a simple Bianchi I universe, since the anisotropic stress term in the stress–energy tensor would be null $\Pi_{ij} = 0$, the shear $\sigma$ would typically decay over time, and anisotropies would be negligible on the local scale, which interests us here. Therefore, to have strong anisotropies at the local scale, we would need a source for the anisotropies, $\Pi_{ij} \neq 0$, and the practical derivation of $\kappa$ becomes model dependent. However, C. Pitrou et al. (2015) also showed that any violation of spatial anisotropy would generate several observables in WL, in particular related to the correlations of the B-mode of shear. With the upcoming data coming from the *Euclid* mission, these observables can be probed (J. Adam et al. 2025). These results will be useful to better constrain the local anisotropies and the impact of possible deviation from a FLRW background on the weak lensing low multipoles.

## 8 CONCLUSION

In this work, we have presented the first combined theoretical, numerical, and observational analysis of the lowest multipoles of the weak lensing convergence field in a $\Lambda$CDM universe. While weak lensing is traditionally explored on intermediate and small angular scales, we have shown that its largest scale components – the dipole, quadrupole, and octopole – carry essential information about the coherent potential gradients generated by the matter distribution within a few hundred Mpc (megaparsecs).

Using exact full-sky (non-Limber) expressions, we demonstrated that the convergence dipole grows with source redshift and rapidly saturates at an amplitude of order $10^{-5}$–$10^{-4}$ by $z_{\rm s} \sim 1$. Through the QUIJOTE simulations, we quantified the dependence of these low multipoles on the observer's environment and found that it was reasonable to expect the low multipoles of the convergence field to be higher than the $\Lambda$CDM expectations for observer within environment such as ours at low redshift, even though the impact of the environment gets smaller as the distances of the observations gets farther away. We found that the amplitudes reconstructed from the 2MRS survey are compatible with the $\Lambda$CDM mean for the dipole and octopole, while the quadrupole was significantly bigger than the $\Lambda$CDM expectation. However, all of these multipoles are compatible with Milky Way-like observers – those residing near a Virgo-like overdensity and exhibiting a CMB-inferred peculiar velocity. Such observers represent only $\sim$1.4 per cent of all particles in the QUIJOTE simulation, confirming that our location is statistically atypical and that average theory predictions do not always apply directly to us.

We then assessed the implications of these results for the long-standing cosmic number count dipole anomaly. By converting the convergence dipole into a number count dipole, we showed that weak lensing produces a small but non-zero effect whose sign and amplitude are controlled by the slope $x$ of the cumulative number counts. For surveys such as CatWISE ($x > 1$), the lensing term enhances the dipole, whereas for NVSS ($x < 1$) it suppresses it. Thus, weak lensing contributes at most a few per cent of the observed anomaly. Our analysis of the 2MRS completeness limit, combined with simple analytic models and numerical tests, shows that structures beyond $z \simeq 0.1$ cannot raise the weak lensing dipole to the level required to reconcile the full discrepancy. Weak lensing is therefore not the dominant origin of the cosmic dipole anomaly, but constitutes a physically unavoidable correction that must be taken into account in future precision analyses.

The main limitation of this study is the depth of the current full-sky redshift catalogues. Although 2MRS provides an excellent map of the local Universe, it is too shallow to capture the full lensing kernel, which extends to $z \gtrsim 0.1$. A definitive determination of the convergence dipole – and of its possible role in wide-angle anomalies – requires deeper, wide-area surveys. The coming decade will provide exactly this. *Euclid* will deliver deep, homogeneous maps of the galaxy distribution and shear field out to $z \sim 2$, enabling direct measurements of the low-$\ell$ convergence modes with unprecedented precision. LSST/Rubin will measure billions of galaxies over half the sky with exquisite photometric depth, providing both improved number count dipoles at high redshift and tomographic access to the full weak lensing kernel. *SPHEREx* (O. Doré et al. 2014), with its full-sky spectrophotometric coverage to $z \sim 1$, will bridge these data sets by supplying accurate photometric redshifts and all-sky maps of large-scale structure. Together, these surveys will allow us to map the matter distribution to far greater distances, construct full light-cone weak lensing reconstructions, and directly measure the convergence dipole, quadrupole, and octopole beyond the reach of 2MRS. They will also make it possible to test whether the small but robust weak lensing contribution identified here plays any role in large-angle cosmological anomalies, and to determine definitively whether the cosmic dipole discrepancy originates from astrophysical selection effects, local structure, relativistic corrections, or a genuine departure from statistical isotropy.

Another limitation of this study is the fact that all analyses have been done within a $\Lambda$CDM framework and FLRW metric. A different cosmological model is expected to yield different results. However, models without an anisotropic stress term within the stress–energy tensor, such as one deriving from a Bianchi I metric, would lead to a decaying geometric shear over time, and therefore negligible anisotropies on the local scale.

Furthermore, it should be noted that the Ellis and Baldwin framework assumes that sources are selected purely by a flux threshold. In many modern surveys, however, catalogues are defined using colour cuts based on multiband photometry. Because the observer's motion Doppler shifts the observed spectrum, spec-tral features can move between photometric bands, potentially changing whether a source satisfies the colour selection. This introduces an additional contribution to the observed number count dipole that is not captured by the original Ellis and Baldwin formalism, as recently discussed by B. Friedman-Shaw et al. (2025). Modelling this effect requires detailed knowledge of the spectral energy distributions of the sources and of the survey selection function. Since this work aims to quantify the weak lensing contribution to the dipole, we have adopted the standard flux-based framework and leave a full treatment of colour-selection effects to forthcoming works.

In summary, the weak lensing convergence low multipoles are not negligible with respect to the higher multipoles, sensitive to the observer's environment, and important in any precision study of the large-angle features in cosmological data. The weak lensing dipole and related modes should be viewed as a foreground/background effect to be modelled or removed. Failing to account for convergence could degrade or bias, for example, PNG or ISW measurements by adding spurious large-scale correlations. In the context of the number count anomalous cosmic dipole, we have shown here explicitly that the lensing contribution can be considered as a source of uncertainty. Upcoming surveys, in particular *SPHEREx*, *Euclid,* and LSST/Rubin, will soon provide the data needed to measure these modes directly going to medium and high redshifts, necessary for a detailed and conclusive result on the extent of the significance of convergence low multipole on large-scale observables of our Universe.

## DATA AVAILABILITY

The data from the QUIJOTE simulations are available at https://quijote-simulations.readthedocs.io/en/latest/, and the 2MRS catalogue is available at http://tdc-www.harvard.edu/2mrs/.

## ACKNOWLEDGEMENTS

We thank Niayesh Afshordi, Reza Ansari, Maciej Bilicki, Johann Cohen-Tanugi, Jacques Colin, Sebastian von Hausegger, Martin Kilbinger, Calum Murray, Nathan Secrest, Jean-Philippe Uzan, and Francesco Villaescusa-Navarro for discussions and helpful suggestions. RM thanks the Beecroft Centre for Theoretical Physics at Oxford University for hospitality.

## APPENDIX A: SPHERICAL HARMONIC ANALYSIS OF AN INTEGRAL ALONG THE LINE OF SIGHT

Consider any field $f$ in the sky $S^2$, which can be written as an integral over the line of sight up to a radius $r_s$ for all directions $\boldsymbol{\theta}$:

$$f(\boldsymbol{\theta}) = \int_0^{r_s} \mathrm{d}r\, w(r, r_s)\, \delta_m(r, \boldsymbol{\theta}), \tag{A1}$$

where $w(r, r_s)$ is any function depending on $r$ and $r_s$ which specifies the relative weight of the matter density along the line of sight in forming the field $f$, and hereafter we write $\delta_m$ for $\delta_m(r, \boldsymbol{\theta}, z = 0)$. The field is observed by an observer at $\boldsymbol{r}_{obs} = 0$.

Including the growth-rate function $D_+(r)$, we can write

$$f(\boldsymbol{\theta}) = \int_0^{r_s} \mathrm{d}r\, w(r, r_s) D_+(r)\, \delta_m. \tag{A2}$$

We aim to write in a simple way the dipole component of this field. For this, we start from the spherical harmonic coefficients $a_{\ell m}$:

$$a_{\ell m} = \int \mathrm{d}\Omega\, Y^*_{\ell m}(\boldsymbol{\theta}) \int_0^{r_s} \mathrm{d}r\, w(r, r_s) D_+(r)\, \delta_m. \tag{A3}$$

In Fourier space, we write

$$\delta_{\rm m} = (2\pi)^{-3} \int {\rm d}^3 k\, {\rm e}^{{\rm i}\boldsymbol{k}\cdot\boldsymbol{r}}\, \delta_{\rm m}(\boldsymbol{k}, z=0). \tag{A4}$$

Using the decomposition of the exponential into spherical harmonics, we have

$$\begin{aligned}
&\int {\rm d}\Omega\, Y^*_{\ell m}(\boldsymbol{\theta})\, \delta_{\rm m}(\boldsymbol{r}, z=0) = \frac{4\pi}{(2\pi)^3} \int {\rm d}\Omega\, Y^*_{\ell m}(\boldsymbol{\theta}) \\
&\times \int {\rm d}^3 k \sum_{\ell', m'} i^{\ell'}\, j_{\ell'}(kr)\, Y^*_{\ell' m'}(\hat{\boldsymbol{k}})\, Y_{\ell' m'}(\boldsymbol{\theta})\, \delta_{\rm m}(\boldsymbol{k}, z=0) \\
&= \frac{1}{2\pi^2} \int {\rm d}^3 k \sum_{\ell', m'} i^{\ell'} \left[ \int {\rm d}\Omega\, Y^*_{\ell m}(\boldsymbol{\theta}) Y_{\ell' m'}(\boldsymbol{\theta}) \right] \\
&\times Y^*_{\ell' m'}(\hat{\boldsymbol{k}})\, \delta_{\rm m}(\boldsymbol{k}, z=0)\, j_{\ell'}(kr) \\
&= \frac{i^\ell}{2\pi^2} \int {\rm d}^3 k\, Y^*_{\ell m}(\hat{\boldsymbol{k}})\, \delta_{\rm m}(\boldsymbol{k}, z=0)\, j_\ell(kr).
\end{aligned} \tag{A5}$$

Therefore,

$$\begin{aligned}
a_{\ell m} &= \int {\rm d}\Omega\, Y^*_{\ell m}(\boldsymbol{\theta}) \int_0^{r_{\rm s}} {\rm d}r\, w(r, r_{\rm s}) D_+(r)\, \delta_{\rm m} \\
&= \frac{i^\ell}{2\pi^2} \int {\rm d}^3 k\, Y^*_{\ell m}(\hat{\boldsymbol{k}})\, \delta_{\rm m}(\boldsymbol{k}, z=0) \\
&\quad \int_0^{r_{\rm s}} {\rm d}r\, w(r, r_{\rm s}) D_+(r)\, j_\ell(kr) \\
&= \frac{i^\ell}{2\pi^2} \int {\rm d}^3 k\, Y^*_{\ell m}(\hat{\boldsymbol{k}})\, \delta_{\rm m}(\boldsymbol{k}, z=0)\, W_\ell(k),
\end{aligned} \tag{A6}$$

where

$$W_\ell(k) = \int_0^{r_{\rm s}} {\rm d}r\, w(r, r_{\rm s}) D_+(r)\, j_\ell(kr). \tag{A7}$$

With this expression, the angular power spectrum is

$$C_\ell = \frac{1}{2\ell+1} \sum_{m=-\ell}^{\ell} |a_{\ell m}|^2. \tag{A8}$$

Inserting $a_{\ell m}$ gives

$$\begin{aligned}
C_\ell &= \frac{1}{2\ell+1} \sum_{m=-\ell}^{\ell} \left| \frac{i^\ell}{2\pi^2} \int {\rm d}^3 k\, Y^*_{\ell m}(\hat{\boldsymbol{k}})\, \delta_{\rm m}(\boldsymbol{k}, z=0)\, W_\ell(k) \right|^2 \\
&= \frac{1}{4\pi^4(2\ell+1)} \sum_{m=-\ell}^{\ell} \iint {\rm d}^3 k\, {\rm d}^3 k'\, Y^*_{\ell m}(\hat{\boldsymbol{k}}) Y^*_{\ell m}(\hat{\boldsymbol{k}}') \\
&\quad \times \delta_{\rm m}(\boldsymbol{k}, z=0) \delta_{\rm m}(\boldsymbol{k}', z=0)\, W_\ell(k)\, W_\ell(k').
\end{aligned} \tag{A9}$$

Using

$$\langle \delta_{\rm m}(\boldsymbol{k}) \delta_{\rm m}(\boldsymbol{k}') \rangle = (2\pi)^3 \delta^{(3)}(\boldsymbol{k} - \boldsymbol{k}') P(k), \tag{A10}$$

we obtain

$$\begin{aligned}
\langle C_\ell \rangle &= \frac{1}{4\pi^4(2\ell+1)} \iint {\rm d}^3 k\, {\rm d}^3 k' \sum_{m=-\ell}^{\ell} Y^*_{\ell m}(\hat{\boldsymbol{k}}) Y^*_{\ell m}(\hat{\boldsymbol{k}}') \\
&\quad \times \langle \delta_{\rm m}(\boldsymbol{k}) \delta_{\rm m}(\boldsymbol{k}') \rangle\, W_\ell(k)\, W_\ell(k') \\
&= \frac{2}{\pi(2\ell+1)} \int {\rm d}^3 k \sum_{m=-\ell}^{\ell} Y^*_{\ell m}(\hat{\boldsymbol{k}}) Y^*_{\ell m}(\hat{\boldsymbol{k}})\, P(k)\, W_\ell(k)^2.
\end{aligned} \tag{A11}$$

The angular integral yields

$$\sum_{m=-\ell}^{\ell} \int {\rm d}\Omega_{\hat{k}}\, Y^*_{\ell m}(\hat{\boldsymbol{k}}) Y^*_{\ell m}(\hat{\boldsymbol{k}}) = 2\ell + 1. \tag{A12}$$

Thus,

$$\begin{cases} \langle C_\ell \rangle \ = \frac{2}{\pi} \int {\rm d}k\, k^2\, P(k, z=0)\, W_\ell(k)^2, \\ W_\ell(k) = \int_0^{r_{\rm s}} {\rm d}r\, w(r, r_{\rm s}) D_+(r)\, j_\ell(kr). \end{cases} \tag{A13}$$

The typical amplitude of the dipole is

$$\langle d^2 \rangle = \frac{9}{4\pi} \langle C_1 \rangle. \tag{A14}$$

## APPENDIX B: CLUSTERING MULTIPOLE

The precedent equation (A13) can be applied to the case of the clustering of galaxies, considering the intrinsic fluctuation of the galaxies in the survey, with $N_\Omega(\boldsymbol{\theta}) = \bar{N}_\Omega(1 + \Delta_{\rm s}(\boldsymbol{\theta}))$, where

$$\begin{cases} \Delta_{\rm s}(\boldsymbol{\theta}) = \int {\rm d}r\, \delta_{\rm s}(r, \boldsymbol{\theta}), \\ \delta_{\rm s}(\boldsymbol{r}) \ = b_{\rm s}(z) n_{\rm s}(z) \delta_{\rm m}(r, \boldsymbol{\theta}), \end{cases} \tag{B1}$$

with $b_{\rm s}(z)$ the bias, and $n_{\rm s}(z) = (N)^{-1}\, {\rm d}N/{\rm d}z$ is the normalized distribution of the sources in the survey. With $w(r, r_{\rm s}) = b_{\rm s}(z) n_{\rm s}(z)$, we have

$$\begin{cases} \langle C^{\rm clust}_\ell \rangle \ = \frac{2}{\pi} \int {\rm d}k\, k^2 P(k, z=0) W_\ell(k)^2, \\ W_\ell(k) \ = \int {\rm d}z\, b_{\rm s}(z) n_{\rm s}(z) D_+(z) j_\ell(kr(z)). \end{cases} \tag{B2}$$

With the redshift distribution of 2MRS, we can obtain, in particular, the expected clustering dipole amplitude:

$$\mathcal{D}_{\rm clust} \approx 2.3 \times 10^{-1}.$$

With a kinematic dipole of the order of $\beta \approx 10^{-3}$, clustering is the dominant effect for 2MRS in the number count dipole.

## APPENDIX C: HIGHER MULTIPOLE OF THE CONVERGENCE, LIMBER APPROXIMATION

Equation (6) gives an exact expression for the convergence expected angular spectrum in ΛCDM; however, it involves a double integral that can be expensive to calculate. The Limber approximation (M. LoVerde & N. Afshordi 2008) is helpful to reduce this kind of expression to a single integral. However, this approximation is only supposed to be valuable at $l \gg 1$. P. Lemos et al. (2017) argued that for the weak lensing shear angular power spectrum, this approximation was good even at low $l$. Here, we show that it is also valuable at low $l$ for the convergence power spectrum. The Limber approximation can be expressed with replacing the spherical Bessel function with a Dirac delta function:

$$j_\ell(kr) \rightarrow \sqrt{\frac{\pi}{2\nu}} \delta(\nu - kr), \tag{C1}$$

where $\nu = \ell + 1/2$. With this, we obtain an expression for the convergence power spectrum:

$$\langle C^\kappa_\ell \rangle \simeq \int_0^{r_{\rm s}} \frac{{\rm d}r}{r} P(k = \nu/r, z=0) \frac{r(r_{\rm s} - r)^2}{r_{\rm s}^2 a(r)^2} + \mathcal{O}(\nu^{-2}). \tag{C2}$$

In Fig. C1, this expression is integrated for $z_{\rm s} = 1$, and we can see that it does indeed converge to the exact solution at high $\ell$, but still gives a correct approximation even at $\ell = 1$. For the dipole, we find $\langle C^\kappa_1 \rangle = 1.65 \times 10^{-8}$ with the exact integration and $1.15 \times 10^{-8}$ with the Limber approximation. However, in Fig. C2, we can see that the Limber approximation overestimates the dominant multipole, and even more so at the bigger redshift.

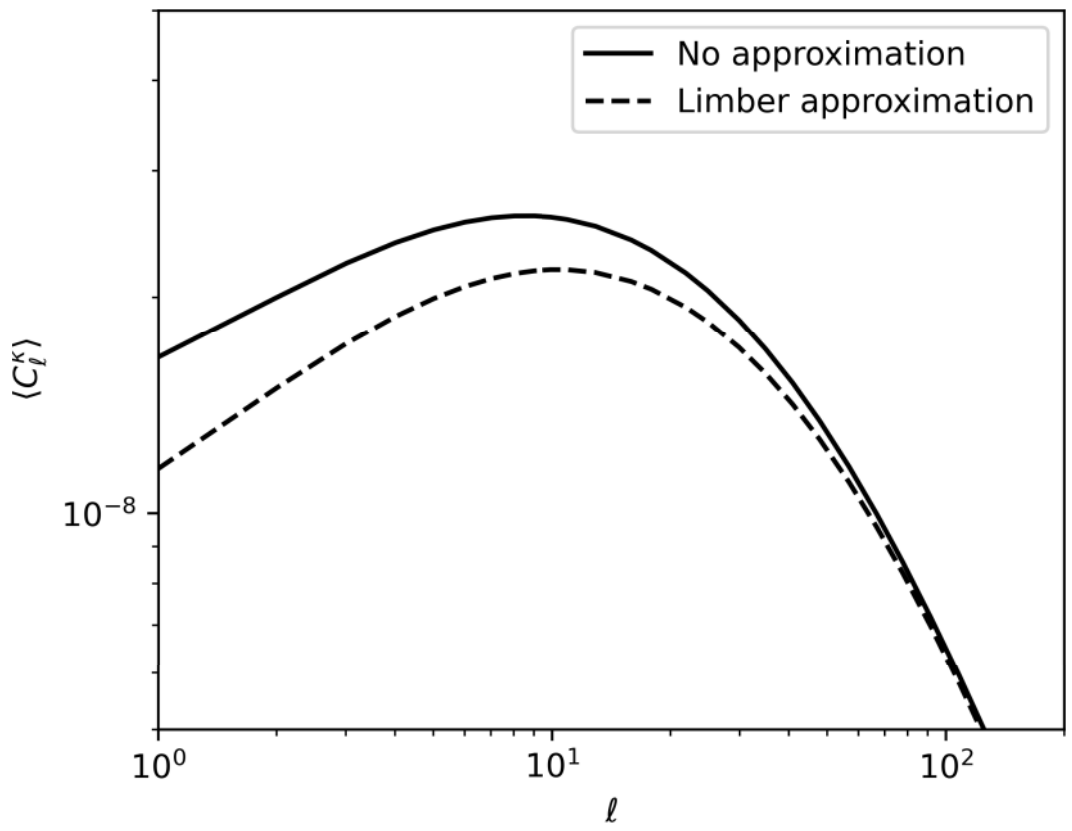

**Figure C1.** Angular power spectrum for the convergence field in $\Lambda$CDM, calculated using the Limber approximation (dashed line), and an exact integration, for sources at $z_s = 1$.

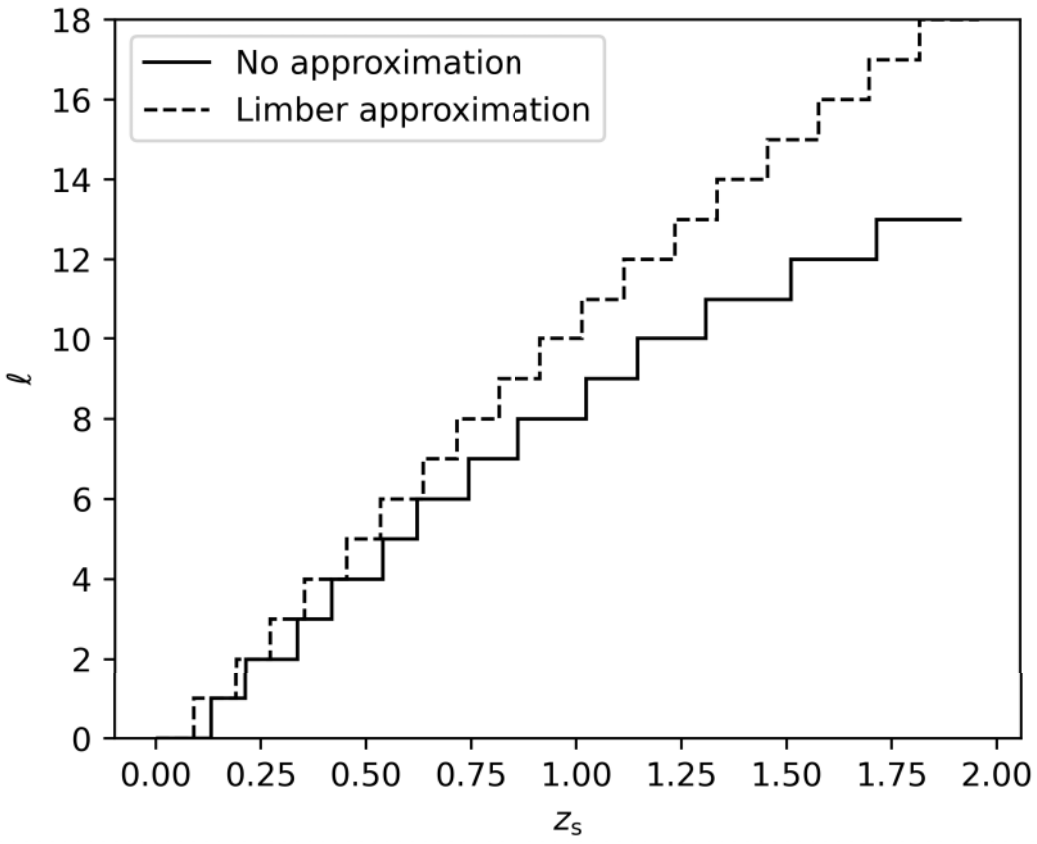

**Figure C2.** Evolution of the dominant multipole up to a redshift $z_s = 2.0$. The exact calculation here is compared to the Limber approximation.

This paper has been typeset from a TEX/LATEX file prepared by the author.